\documentclass[format=acmsmall, review=false, screen=true]{acmart}
\usepackage{booktabs} 
\usepackage{graphicx}
\usepackage{subfigure}
\usepackage{array}
\usepackage[ruled]{algorithm2e} 

\SetAlFnt{\small}
\SetAlCapFnt{\small}
\SetAlCapNameFnt{\small}
\SetAlCapHSkip{0pt}
\IncMargin{-\parindent}

\acmVolume{x}
\acmNumber{x}
\acmArticle{00}
\acmYear{20xx}
\acmMonth{0}

\renewcommand\footnotetextcopyrightpermission[1]{} 
\acmDOI{xxxxx.xxxxxx}

\received{Month Year}
\received[revised]{Month Year}
\received[accepted]{Month Year}

\begin{document}
\title[GARDENIA]{GARDENIA: A Domain-specific Benchmark Suite for Next-generation Accelerators}  
\author{Zhen Xu}
\author{Xuhao Chen}
\author{Jie Shen}
\author{Yang Zhang}
\author{Cheng Chen}
\author{Canqun Yang}
\affiliation{%
  \institution{National University of Defense Technology}
  \department{College of Computer}
  \city{Changsha}
  \state{Hunan}
  \postcode{410073}
  \country{China}
}

\begin{abstract}
This paper presents the Graph Analytics Repository for Designing 
Next-generation Accelerators (GARDENIA), a benchmark suite for 
studying irregular algorithms on massively parallel accelerators.
Applications with limited control and data irregularity are the main focus of existing generic benchmarks for accelerators, while available graph analytics benchmarks do not apply 
state-of-the-art algorithms and/or optimization techniques. GARDENIA 
includes emerging irregular applications in big-data and machine learning 
domains which mimic massively multithreaded commercial programs running 
on modern large-scale datacenters. Our characterization shows that GARDENIA 
exhibits irregular microarchitectural behavior which is quite different 
from structured workloads and straightforward-implemented graph benchmarks. 
\end{abstract}



\thanks{This work is supported by the National Science Foundation of China,
  Authors' addresses: Zhen Xu, Xuhao Chen, Jie Shen, Yang Zhang, Cheng Chen, Canqun Yang, 
	National University of Defense Technology, 109 Deya Rd, Kaifu District, Changsha, Hunan 410073, China.
}

\maketitle
\renewcommand{\shortauthors}{Z. Xu et al.}
\section{Introduction}
With the rise of big-data analysis and machine learning, graph analytics 
applications have become increasingly important and pervasive. 
For these frequently executed algorithms, dedicated hardware accelerators
~\cite{Ahn,Ozdal,Ham} are an energy-efficient avenue to high performance.
Meanwhile, general purpose accelerators (e.g., GPUs and MICs) which have been
widely deployed in many HPC systems and datacenters, are also applied to 
speedup applications in these areas, such as graph analytics~\cite{nvGRAPH},
machine learning~\cite{cuDNN} and sparse linear algebra~\cite{CUSPARSE}.
However, these applications exhibit irregular runtime behavior which is 
quite different from that of conventional well-studied HPC applications.
Therefore, it is important to set up a standard benchmark suite for
architecture researchers to get deep understanding on these applications.

The goal of GARDENIA\footnote{The source code can be found at
\href{https://github.com/chenxuhao/gardenia}{github.com/chenxuhao/gardenia}}
is to develop an irregular benchmark suite for future 
accelerator architecture research, including both general-purpose and 
domain-specific accelerator design. For general-purpose accelerators, GARDENIA
is an important complement for generic benchmark suites (e.g.,Rodinia and 
Parboil) which have limited irregularity, and architectural support for 
irregular algorithms can be exploited by running the benchmarks and locating
the performance bottlenecks. For domain-specific accelerators, GARDENIA 
can be used as a workload representative of real-world graph analytics 
workloads, and in this way it enables the  hardware specialization for these algorithms. 
Besides, GARDENIA benchmarks can be also used by algorithm, library 
and compiler designers as reference implementations for comparison.

As far as we know, GARDENIA is the first graph analytics benchmark
suite specifically targeting accelerator architecture reserach. It is 
different from previous GPU graph analytics benchmark suites in many ways:
1) The benchmark implementations incorporate \textit{state-of-the-art} 
optimization techniques for massively parallel accelerators;
2) The input graphs are selected appropriately for our target platforms, 
in terms of graph size, desity and topology;
3) The workloads are representatives of popular graph analytics, machine 
learning and sparse linear algebra applications running on modern datacenters.

This paper makes the following three major contributions:
\begin{itemize}
\item It unveils the limitations of previously available benchmark suites
and identifies the reason for which they are not good candidates to evaluate future accelerators.

\item We present GARDENIA, a domain-specific benchmark suite for accelerators 
that provides irregularity, diversity and state-of-the-art techniques.

\item We characterize GARDENIA, illustrate its microarchitecture behavior,
and analyze the performance bottlenecks to provide insights on how to design
next-generation accelerators with energy-efficiency for irregular workloads.
\end{itemize}

The rest of the paper is organized as follows:
Section~\ref{sect:motivation} explains the motivation of this work.
Section~\ref{sect:design} describes the design of our benchmark suite.
Then we characterize the benchmarks in Section~\ref{sect:evaluation}.
And section~\ref{sect:conclusion} concludes.

\section{Motivation}\label{sect:motivation}
This work aims to develop an irregular graph benchmark suite that can help designing 
next-generation accelerators capable of accelerating real-world irregular 
(not only HPC) applications on modern datacenters. In section 2.1, we show what features such a benchmark suite should equip. Then in section 2.2, we discuss how existing benchmark suites cannot fulfill these conditions.

\subsection{Benchmark Requirements}
A benchmark suite of real-world irregular workloads on data parallel 
accelerators needs to meet the following requirements:

\textbf{Massive Parallelism and Accelerator Coverage}
Massively parallel accelerators, such as GPUs and MICs (Xeon Phi coprocessors)
have been already widely deployed in today's high performance servers and 
datacenters. Essential big-data analysis and machine learning engines that 
drive many important applications are highly dependent on accelerators to 
achieve required performance. 
Future accelerators' main feature is to provide tremendous performance 
using extreme throughput and memory bandwidth. Meanwhile, 
to take full advantage of the hardware, workloads should be well paralleled.

\textbf{Workload Irregularity}
Regular applications have been intensively investigated on large-scale parallel 
systems in the past decade. Many scientific and commercial applications, 
have been successfully mapped onto accelerators. 
Despite this progress, data-parallel accelerators (DPAs) continue to be confined 
to structured parallelism. While structured parallelism maps directly to DPAs, 
much more unstructured applications in real-world cannot simply take advantage 
of them. GARDENIA is designed to represent irregular graph applications, and can be used to guide the design of future domain-specific accelerators. 

\textbf{State-of-the-Art Optimization Techniques}
Researchers have made great efforts to map general purpose applications onto
GPUs during the past decade, including parallel algorithm innovations and 
optimization techniques. A benchmark suite should include rising optimizations to fully represent real-world applications, which are rarely straight-forward implementations.
This is important because straightforward-implemented workloads may exhibit
quite different microarchitectural behaviors compared to optimized ones, 
which are unqualified to represent real-world applications and may mislead 
architecture design.

\textbf{Workload and Dataset Diversity}
With the rise of heterogeneous computing, real-world workloads are 
increasingly diverse. They are written in different parallelization models, 
run on a variety of computing platforms, and cover a wide range of 
application areas. 
Besides, the behaviors of these irregular applications change 
dramatically with different types of input datasets. To cover 
different program characteristics and behaviors, a diverse 
collection of both workloads and input datasets is required, 
so that architects can thoroughly understand the application 
behaviors and make reasonable trade-offs when designing an accelerator.

\subsection{Existing Benchmark Suites and their Limitations}
We further discuss the drawbacks of previously available benchmark suites. Table \ref{table:suite_comparison} shows 
the comparison between existing benchmark suites and GARDENIA. 

\textbf{Generic Benchmark Suites for Multicore CPUs.}
\textit{PARSEC}~\cite{PARSEC} is an application repository for shared-memory computers. It is composed of several shared-memory 
multithreaded workloads from emerging programs. This suite only 
includes CPU programs (e.g. Pthreads, OpenMP and TBB programs) 
and is not specifically designed for irregular workloads.

\textbf{Generic Benchmark Suites for Accelerators.} 
\textit{Parboil}~\cite{Parboil} and \textit{Rodinia}~\cite{Rodinia}
are widely used GPU benchmark suites. They provide CUDA, OpenCL and 
OpenMP implementations. Despite the popularity, most of their benchmarks 
are structured kernels from HPC applications with limited control and 
memory irregularity. Although they do include some irregular workloads 
such as BFS, their implementations are quite out-of-date. Thus, they are 
not good candidates for studying emerging irregular applications on accelerators.

\begin{table}[htb]
	\begin{center}
		\scriptsize
		\begin{tabular}{c c c c c c}
			\hline
			\hline
			\textbf{} & \parbox{1.5cm}{\centering \textbf{No. of workloads\\}} & 
			\parbox{1.5cm}{\centering \textbf{Accelerator\\ oriented}} & 
			\parbox{1.5cm}{\centering \textbf{Irregular}} & 
			\parbox{1.5cm}{\centering \textbf{Diverse}} & 
			\parbox{1.8cm}{\centering \textbf{up-to-date}} \\
			\hline
			GARDENIA & 9 & $\surd$ & $\surd$ & $\surd$ & $\surd$\\
			\hline
			PARSEC & 12 &  $\times$ & $\times$ & $\surd$ & $\surd$\\
			\hline
			Parboil & 11 & $\surd$ & $\times$ & $\surd$ & $\surd$\\
			\hline
			Rodinia & 23 & $\surd$ & $\times$ & $\surd$ & $\surd$\\
			\hline
			GAPBS & 6 &  $\times$ & $\surd$ & $\times$ & $\surd$\\
			\hline
			Pannotia & 8 & $\surd$ & $\surd$ & $\surd$ & $\times$\\
			\hline
			GraphBIG & 6 & $\surd$ & $\surd$ & $\surd$ & $\times$\\
			\hline
			LonestarGPU & 7 & $\surd$ & $\surd$ & $\times$ & $\surd$\\
			\hline
			\hline
			\\
		\end{tabular}
		\caption{Comparison between GARDENIA and Previously Available Benchmark Suites.
			"No. of wrokloads" shows how many workloads are included; 
			"Accelerator oriented" shows if the benchmark suite is built targeting 
			accelerators; "Irregular" shows if the suite focuses on irregular 
			workloads; "Diverse" shows if the suite includes diverse 
			workloads and datasets; "up-to-date" shows if the suite 
			employs up-to-date optimization techniques.
		}
		\label{table:suite_comparison}
		\vspace{-0.5cm}
	\end{center}
\end{table}

\textbf{Graph Benchmarks for CPUs.} 
\textit{GAPBS}~\cite{GAPBS} is a graph processing 
benchmark suite on CPUs. It is a portable high-performance graph baseline using only OpenMP for parallelism. Their implementations are 
representatives of up-to-date performance on multicore CPUs.
However, GAPBS does not include any implementation for accelerators.
Besides, it only includes six typical graph benchmarks, which is not
diverse enough to represent emerging irregular applications on accelerators.

\textbf{Graph Benchmarks for GPUs.} 
\textit{Pannotia}~\cite{Pannotia} assembles a set of graph algorithms
implemented in OpenCL, and the irregularities of these kernels are 
characterized on GPUs. However, due to limited knowledge on how to optimize 
graph algorithms on GPUs at that time, no specific optimization is applied.
We will show that workloads which lack of state-of-the-art optimization
techniques behaves quite differently from optimized ones in Section~\ref{sect:evaluation}.
\textit{GraphBIG}~\cite{GraphBIG} includes workloads for both CPU and GPU sides, using OpenMP and CUDA respectively. Since it mainly focuses on 
CPU versions, most of its GPU workloads are also straightforward implemented,
although some of the GPU workloads achieve substantial speedup.
\textit{Lonestargpu}~\cite{Lonestargpu} includes a couple of irregular CUDA 
benchmarks with advanced optimizations, but it does not focus on emerging big-data
and mechine learning applications. Thus it is not an appropriate candidate for 
designing next-generation accelerators for these domains. Besides, the datasets 
it provides are not diverse enough for deeply understanding graph algorithms. 

\textbf{Graph Processing Frameworks.} 
Many parallel CPU (Pregel~\cite{Pregel}, GraphMat~\cite{GraphMat}, Ligra~\cite{Ligra}, 
Graphlab~\cite{Graphlab,DistributedGraphLab}, GraphReduce~\cite{GraphReduce}) 
and GPU (Gunrock~\cite{Gunrock}, Medusa~\cite{Medusa}, CuSha~\cite{CuSha}) graph 
processing frameworks are developed to provide programmability and high 
performance simultaneously. Efforts have been made to generalize the optimization techniques that 
are previously proposed in specific graph algorithms. These frameworks provide 
workloads that employ state-of-the-art techniques, but they can not be efficiently 
used for architecture research due to the complexity of their infrastructures. Note 
that they are not benchmarks suites, but they provide us the insight that guides 
our design. We applies the key optimizations in these frameworks to our workloads, 
making sure that our benchmark suite represents real-world applications.\\

Our GARDENIA benchmark suite tackles the limitations of existing benchmark suites, applies the state-of-the-art optimization techniques, and covers diverse irregular workloads and datasets. Thus, GARDENIA is a good benchmark candidate for architecture research. 

\section{The GARDENIA Benchmark Suite}\label{sect:design}
GARDENIA is intended to provide a graph workload collection 
that represents important irregular applications running on the 
accelerators of modern datacenters for big-data analysis and machine learning. And in this way it can help graph processing research and accelerator design by standardizing evaluations.
GARDENIA fully fulfill the demands highlighted in Section~\ref{sect:motivation}:

\begin{itemize}
\item Every workload is parallelized using OpenMP for 
multicore CPUs, CUDA for GPUs and OpenMP target for MICs 
to exploit massive parallelism.

\item GARDENIA is not designed for regular HPC programs which have 
been well studied on accelerators in the last decade. It focuses on 
emerging irregular workloads that mimic actual machine-learning and big-data applications. 

\item All the benchmarks apply state-of-the-art optimization techniques,
i.e., the benchmarks are reasonably optimized to get substantial
speedups on GPUs and MICs.

\item The workloads as well as their input datasets are multitudinous and are carefully selected 
from various application domains.

\end{itemize}

\subsection{Optimization Techniques}
We apply the state-of-the-art algorithm innovations and optimization 
techniques that have been widely employed in academic and industry 
libraries ~\cite{Gunrock,CUSPARSE,CUSP}. Benchmarks are neither
over-optimized nor unoptimized (i.e. straightforward-implemented) 
for the target accelerator. Techniques bound to specific 
architectures are not included in our suite. 

\textbf{Mapping Strategies.} 
Basically there are two parallelism strategies for graph algorithms:
\textit{quadratic} mapping and \textit{linear} mapping~\cite{DVT}.
Quadratic mapping checks every vertex of the graph in each iteration,
and depending 
on if the vertex needs to be processed a thread may process its vertex or stay idle ~\cite{Merrill}.
Quadratic mapping is intuitive and used for sequential-movement
benchmarks. 
By contrast, a frontier queue is maintained in linear mapping stargegy to hold the to be processed vertices during each iteration.
Then corresponding threads are initialized and the number of threads is due to amount of vertices to be processed.
Same number of to be processed vertices are distributed to each thread and therefore all threads are working till they terminate at the same time.~\cite{Merrill}. In this way, 
linear mapping is commonly more efficient than quadratic
mapping, but extra overhead arises from maintaining the frontier. 
We use linear mapping for traversal-based benchmarks 
(see section 3.2 for details).

\textbf{Load Balancing.}
Load imbalance problem is one of the key issues for irregular algorithms, 
and is particularly worse for scale-free (power-law) graph datasets.
Hong~\emph{et~al.}~\cite{Hong} proposed a warp-execution method for BFS 
to map warps instead of threads to vertices. This scheme is applied
in most of the GARDENIA benchmarks. Based on Hong's scheme, 
Merrill~\emph{et~al.}~\cite{Merrill} proposed a multi-level load balancing 
scheme. 
According to the size of a vertex's neighbor list, the workload of a vertex may be distributed to a thread, a warp or a thread block.
This strategy turns out to be quite efficient on GPUs, and is employed in 
traversal-based benchmarks in GARDENIA.

\textbf{Push vs. Pull}
Beamer~\emph{et~al.}~\cite{Beamer} proposed direction-optimizing BFS which
uses two different methods to drive BFS: \texttt{push} and \texttt{pull}. 
The \texttt{push} method detects the status of vertices in current frontier and scatters 
them to their outgoing neighbors to create new frontier. The \texttt{push} 
method is intuitive and commonly used. Conversely, the \texttt{pull} method 
checks unvisited vertices whose parents are in the current frontier and gathers
status from the incoming neighbors. The \texttt{pull} method is beneficial if
the unvisited vertices' number is less than the frontier size
~\cite{Gunrock}. 

\textbf{Reordering Queue.} 
The order of vertices in the frontier queue decides the computation order, memory
access order and load balance, and therefore affects performance. By ochestrating 
the order of vertices for processing, many graph algorithms can reduce overall 
computation workload or memory access latency. For example, we use the delta-stepping 
~\cite{delta} implementation for OpenMP SSSP on CPUs and MICs. This approach organizes 
vertices in the output frontier into ``bin'' or ``buckets'' by their distances to the
source vertex and those vertices with smaller distances get to be processed first. 

\textbf{Other Techniques.} We also include architectural optimization techniques
to take advantages of the underlying hardware. For example, we use the \textit{texture 
cache} to hold the read-only data, i.e. the $x$ vector in \texttt{spmv}, and thus 
the read-only data is forced to be cached in the L1 read-only cache which has 
much shorter latency than the off-chip DRAM. For MIC benchmarks, we enable 
\textit{vectorization} to make full use of the compute capability. We also use 
\textit{bit operations} to reduce computation and storage overhead, i.e., in linear 
mapping algorithms on the frontier queue, the current frontier is converted 
into a bitmap of vertices, and we get to know whether a vertex is active only 
by checking if they are valid in the bitmap. For many CUDA implementations,
\textit{kernel fusion} combines multiple kernels into a single one, and thus 
can keep the entire program on the accelerator. This is beneficial because
it leverages producer-consumer locality between adjacent kernels~\cite{Gunrock}.

\subsection{Diverse Irregular Workloads}
The GARDENIA suite currently includes the following 9 workloads(many more on the way to release):

\textbf{Breadth-First Search (BFS)} is a key graph primitive which
traverses a graph in breadth-first order~\cite{Merrill}. It starts at the tree root and explores the neighbor nodes first before moving to the next level neighbours.
We implement it using the linear mapping strategy along with Merrill's load balancing technique.

\textbf{Single-Source Shortest Paths (SSSP)} finds the shortest path 
from a specific source vertex to every other reachable vertices  in a directed graph~\cite{SSSP}. We implement 
delta-stepping algorithm for CPUs and Bellman-Ford algorithm for GPUs.
SSSP is a traversal-based algorithm and we apply similar techniques as in BFS.

\textbf{Betweenness Centrality (BC)} 
is a measure of centrality in a graph based on shortest paths. For every pair of vertices in a connected graph there exists at least one shortest path between them. The betweenness centrality for each vertex is the number of these shortest paths that pass through the vertex.
BC is also traversal-based.

\textbf{PageRank (PR)} ranks a website based on the score of its
neighboring sites (i.e. the sites that link to it)~\cite{PageRank}. 
Each vertex is given a initial score at the beginning. At each 
iteration, it scans each vertex and updates its score using the 
its neighbors' scores and degrees. It iterates until the score 
change is small enough. Unlike above traversal-based benchmarks, 
PR has \texttt{sequential-movement} access pattern and so do all 
the following benchmarks.

\textbf{Connected Components (CC)} 
All the vertices of a conncected component are assigned a unique label~\cite{BeamerAP15}. 
In each iteration, the label is propagated from a smaller-id vertex 
to its larger-id neighbors. The algorithm iterates until convergence 
(no propogation happens). 

\textbf{Triangle Counting (TC)} counts how many triangles there are in 
an undirected graph~\cite{IMP}. It is essential in graph statistics 
applications (e.g. clustering coefficients). To find triangles, it computes the neighbor list
intersection of 
each vertex and its neighbor. 

\textbf{Stochastic Gradient Descent (SGD)} 
is a stochastic approximation of the gradient descent optimization.
It decomposes the $ratings$ matrix into two feature vectors ($users$ 
and $items$), and it uses the dot product of $users$ and $items$ to make a rating prediction. It iterates until the error between prediction 
and the actual rating is small enough.

\textbf{Sparse Matrix-Vector Multiplication (SpMV)} is one of the most heavily used primitives in scitific computations~\cite{IMP}. 
It identifies the elements in a matrix that are non-zero and multiplies them in a vector. 

\textbf{Symmetric Gauss-Seidel smoother (SymGS)} is an important 
primitive in High Performance Conjugate Gradients (HPCG)~\cite{CUSP}. 
It carries out a forward and backward triangular solve whose
operation is similar to SpMV but in a data-dependent order.
We use vertex coloring~\cite{VC} to find parallelism and
arrange the order of tasks.

\subsection{Diverse Input Datasets}
A major characteristic of irregular applications is the input dependency. 
Research has shown that graph properties substantially affect performance~\cite{GAPBS}.
Therefore, we should characterize graph workloads not only by the algorithms,
but also by the properties of graph instances used. 

\begin{table}[]
	\small
	\centering
	\begin{tabular}{c c c c c}
		\hline
		\hline
		\bf{Graph} & \bf{\# Vertices} & \bf{\# Edges} & \bf{Avg. deg.} & \bf{Description}\\
		\hline
		\texttt{cage14} & 1.5M & 27M & 18 &  DNA electrophoresis\\
		\hline
		\texttt{Freescale} & 3.0M & 23M & 7.7 & Circuit simulation\\
		\hline
		\texttt{Flickr} & 2.3M & 33M & 14.4 & Social relationship inside flickr.com\\
		\hline
		\texttt{Web-Google} & 0.9M & 5.1M & 5.8 & Webpage graph from Google\\
		\hline
		\texttt{Youtube} & 1.1M & 4.9M & 4.3 & Relationships of Youtube users\\
		\hline
		\texttt{LiveJ} & 4.8M & 69M & 14.2 & LiveJournal friendship  network\\
		\hline
		\texttt{Soc-Pokec} & 1.6M & 31M & 18.8 & Pokec user relationship\\
		\hline
		\texttt{Wikipedia} & 3.1M & 39M & 12.5 & Wikipedia page link graph\\
		\hline
		\hline
	\end{tabular}
	\caption{Suite of benchmark datasets}
	\label{table:bench_graphs}
\end{table}

\textbf{Graph Size}
Due to the rapid increase of data size in modern datacenters, it has become
problematic to manage and manipulate the large volume of data with relatively
limited hardware resources. Our selected datasets should include large enough
graph instances for big-data analysis. On the other hand, to support 
architecture research, small but representative datasets are also needed 
to guarantee that the simulation finishes in reasonable time. 

\textbf{Graph Density/Sparsity} 
The average degree of vertices is usually used to represent the density of a graph.
This parameter affects the computation intensity, irregularity, 
and the amount of locality exists in graph algorithms. A denser graph 
is likely to have more data reuse since vertices may be visited or updated 
more times. High density also means more regular memory access behavior. However, 
denser graphs have more edges, which implies larger working set. 
In most case real-world graphs are sparse, which makes graph analytics irregular.

\textbf{Graph Topology}
In ~\cite{GAPBS}, graphs are divided into two categories in terms 
of topology: meshes and social networks.  
Meshes usually have ralatively higher diameter and lower degree distribution because they are usually obtained from physical world, such as electric transmission network. 
By contrast, social 
networks are non-spatial, and they usually have a power-law degree distribution (scale-free).
Due to degrees
(number of neighbors) of different vertices may vary much in a social network,
load balancing is a major issue for parallel execution~\cite{GAPBS}. 

In summary, we select our datasets from the UF Sparse 
Matrix Collection~\cite{davis2011university}, the SNAP dataset Collection~\cite{SNAP}, and the 
Koblenz Network Collection~\cite{kunegis2013konect} which are all publicly available. 
Size, density, topology and application domains vary among these graphs. 
For example, \texttt{twitter} is a large ``social network'' graph that 
can be used to evaluate real machines, while \texttt{road} is a relatively 
small-sized graph with ``mesh'' topology that can be used in simulation.
We list part of the datatsets used for the following evaluation in Table 
\ref{table:bench_graphs}.

\section{Evaluation}\label{sect:evaluation}

In this section, we evaluate GARDENIA in terms of workload irregularity, optimization 
effects, workload/dataset diversity, and accelerator coverage, showing the necessity of developing this 
irregular workload benchmark suite.

\subsection{Experimental Setup}
We test and compare 3 implementations in this paper including: 
(1) \texttt{Serial}: Serial implementation on CPU,
(2) \texttt{CUDA}: CUDA implementation on GPU.
(3) \texttt{MIC}: OpenMP target implementation on MIC(Xeon Phi).

\texttt{Serial} experiments are executed on dual socket Intel Xeon E5-2620 2.10 GHz 6-core CPU. We perform the \texttt{CUDA} experiments on NVIDIA K40m GPU with CUDA Toolkit 8.0
release. For the \texttt{MIC} experiments, we launch 228 threads on MIC, 4 threads on each of 57 cores. Option \texttt{-O3} and \texttt{-arch=sm\_35} are used respectively for \texttt{gcc} and \texttt{nvcc} to compile the \texttt{CUDA} implementations.

All the benchmarks are executed for 10 times and the average execution time is collected. We only focus on the computation part of each program's timing. For the \texttt{CUDA} implementation test, we exclude the data in/out transfer time, which is about 10\%-15\% of the entire execution time.

\subsection{Workload Irregularity}
Branch divergence and memory divergence are two critical factors that reflect workload 
irregularity. We use two metrics, BDR and MDS, to measure them. BDR (Branch 
Divergence Rate) is the average ratio of inactive threads per warp, which reflects the 
amount of branch divergence\cite{nai2015graphbig}. MDS (Memory Divergence Scale) is the average number 
of executed instruction's replays due to memory issues. The replay will be repeated until all requested data are ready, and therefore MDS may be greater than 1.

\begin{figure}[!ht]
\subfigure[BDR]{
	\label{fig:irregularity:bdr}
	\includegraphics[width=0.45\textwidth]{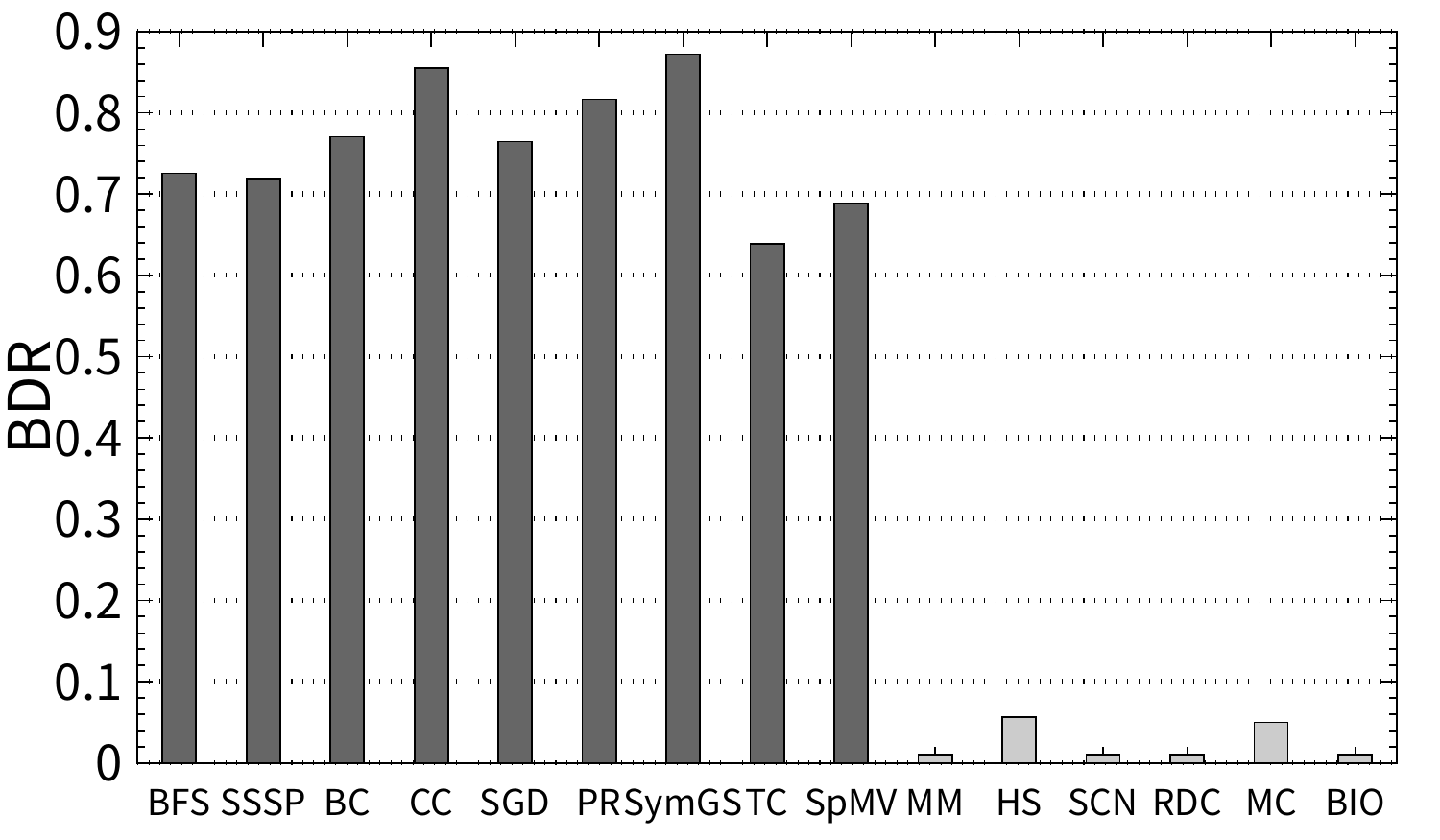}
	}
\subfigure[MDS]{
	\label{fig:irregularity:mds}
	\includegraphics[width=0.45\textwidth]{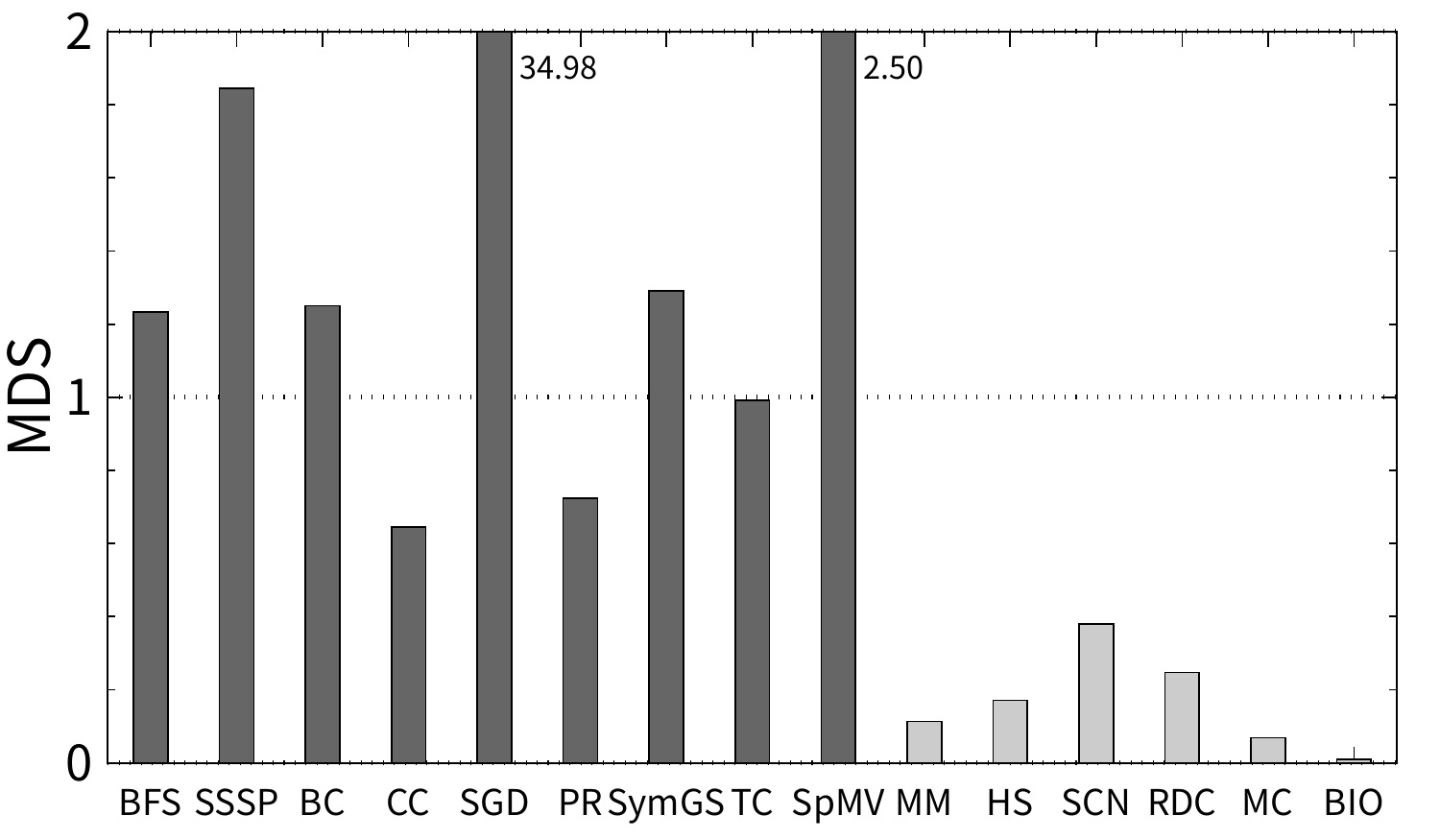}
}
  \vspace{-0.3cm}
  \caption{The difference between irregular and regular workloads.}
  \label{fig:irregularity}
\end{figure}

We select 9 workloads from GARDENIA as irregular workloads along with 6 workloads from  
NVIDIA SDK \cite{nvidia2013computing} and Rodinia \cite{che2009rodinia} as regular ones for comparison. The regular workloads covers various algorithms in specific application domains: Matrix Multiplication (MM) is from the dense liner algebra area; 
Hotspot (HS) performs physical simulation; Scan (SCN) and Reduction(RDC) are widely used parallel computing primitives; 
MonteCarlo (MC) and Binomial Options (BiO) are financial applications.
We run all the workloads with the same input dataset.

Figure~\ref{fig:irregularity} 
shows that GARDENIA workloads have much higher BDR and MDS compared to the 
regular ones, which demonstrates that irregular 
workloads in real-world big-data applications have high branch and memory diversities. The high irregularity
makes GARDENIA workloads behave significantly different from those structured 
benchmarks found in previous generic suites, which implies the necessity of
constructing an irregular benchmark suite like GARDENIA.


\subsection{Optimization Effects}
We employ state-of-the-art optimization techniques for each GARDENIA workload. 
This is essential because straightforward implementations may show largely different 
microarchitectural behaviors and mislead architecture design. We compare GARDENIA and straightforward CUDA implementations to show our optimization effects.


\begin{figure}[!ht]
	\subfigure[SIMT utilization]{
		\label{fig:optimization:simt}
		\includegraphics[width=0.45\textwidth]{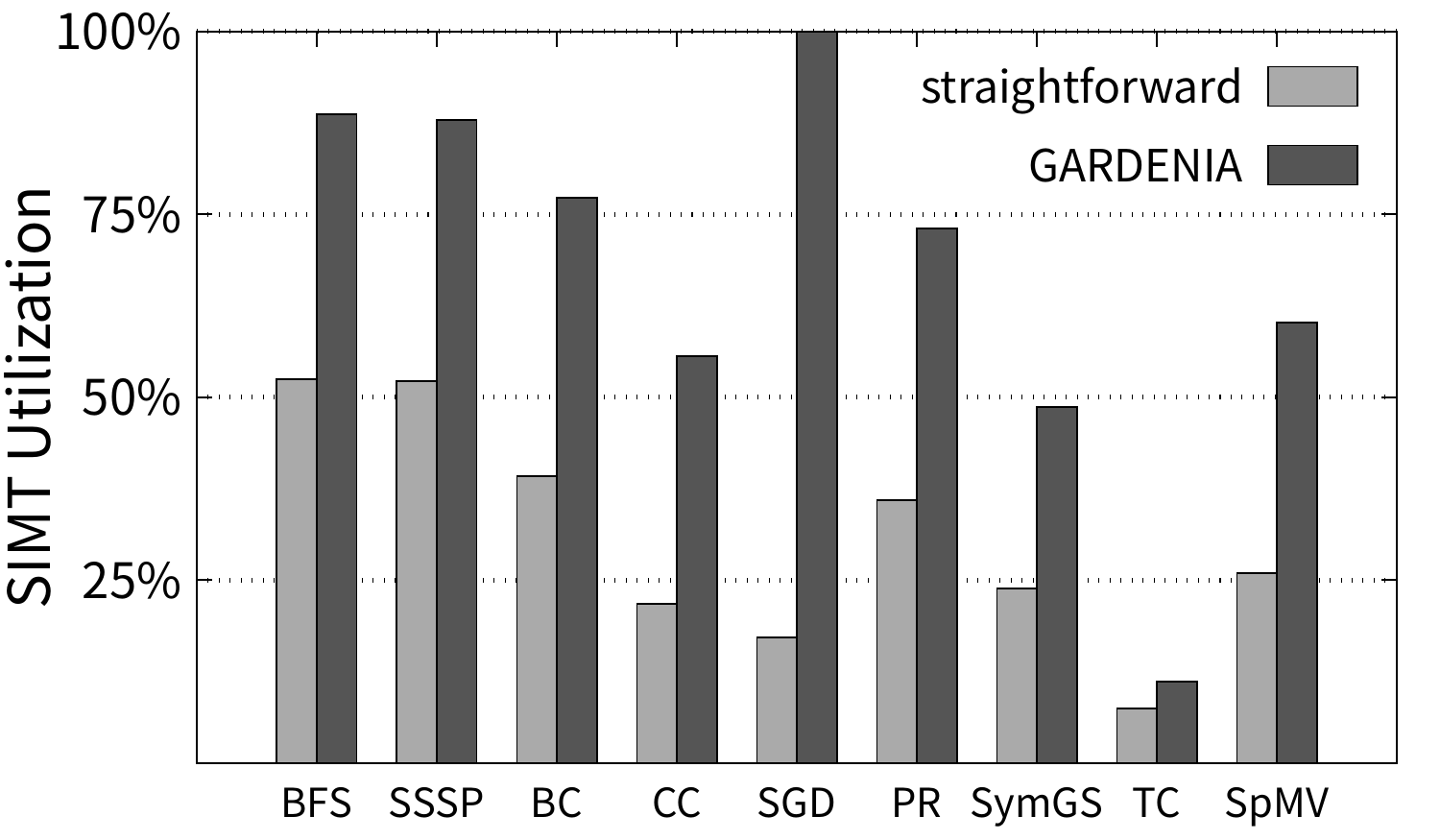}
	}
	\subfigure[IPC]{
		\label{fig:optimization:ipc}
		\includegraphics[width=0.45\textwidth]{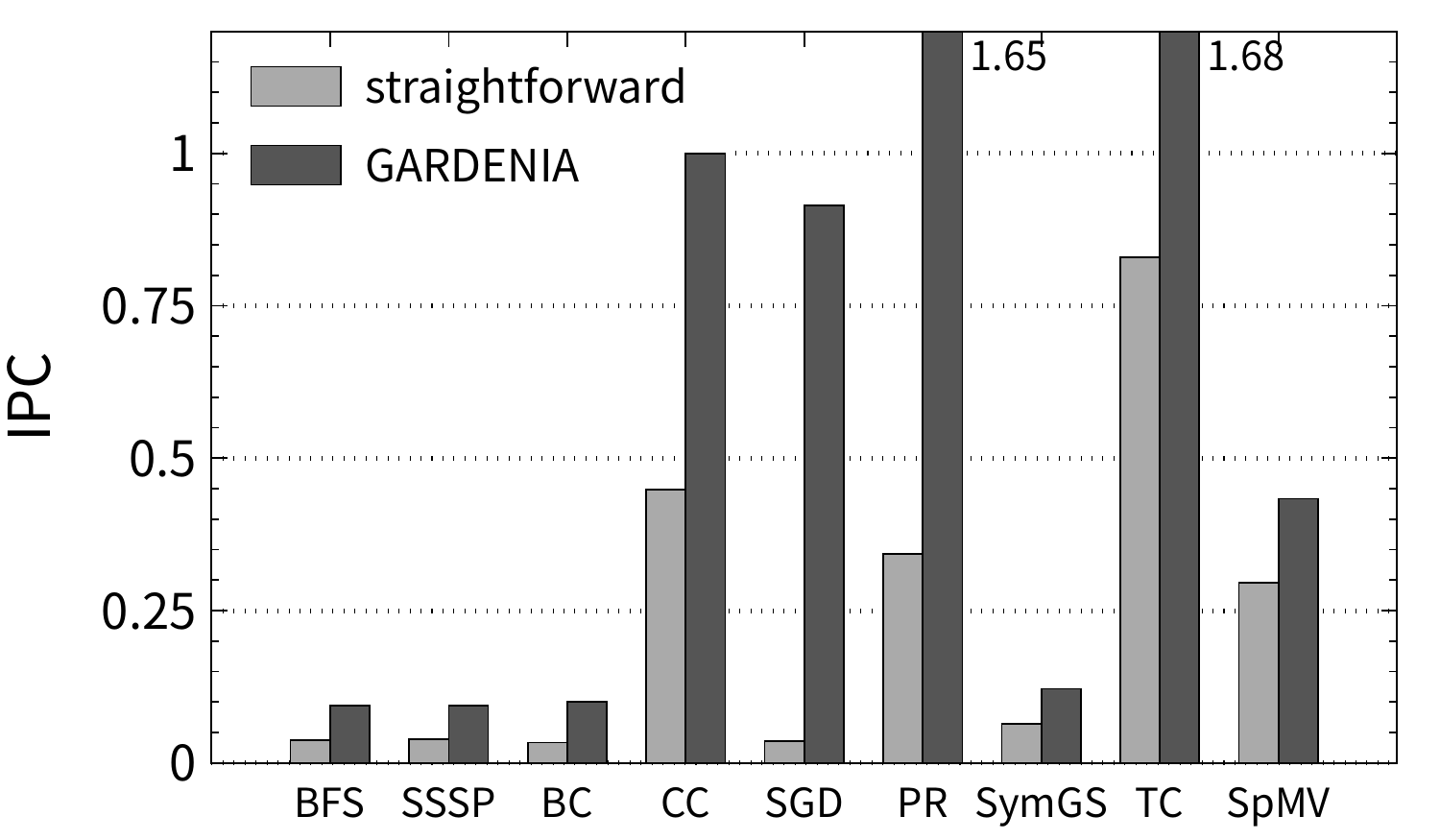}
	}
	\vspace{-0.3cm}
	\caption{Optimization effects measured by the SIMT utilization and IPC.}
	\label{fig:optimization}
\end{figure}

Figure~\ref{fig:optimization} compares the SIMT utilization and IPC(Instructions per cycle) of the \textit{straightforward} 
implementations (similar to those provided in Pannotia \cite{Pannotia} and GraphBIG \cite{nai2015graphbig}) and the 
\textit{GARDENIA} implementations with optimizations applied.  We observe that the SIMT 
utilization as long as the executed IPC of the GPU are largely increased by applying optimization, leading up to 3.8 $\times$ performance improvement. This huge difference 
indicates that optimization techniques can substantially change program behavior, 
so it is necessary to include up-to-date techniques in benchmark suite so as to closely mimic the 
behavior of real-world applications which are most likely built using these optimizations.

\subsection{Workload and Dataset Diversity}

To represent emerging irregular workloads for next-generation accelerators, 
GARDENIA selects representative workloads from various application domains. 
Furthermore, as graph processing performance depends not only on the workload 
but also on the dataset to run, a careful collection of graph datasets is of necessity. We show this with the CUDA implementations of 8 GARDENIA workloads with 5 different datasets. SGD uses quite different datasets as others, so we exclude SGD from this comparison.

\begin{figure}[!ht]
\centering
\includegraphics[width=0.90\textwidth]{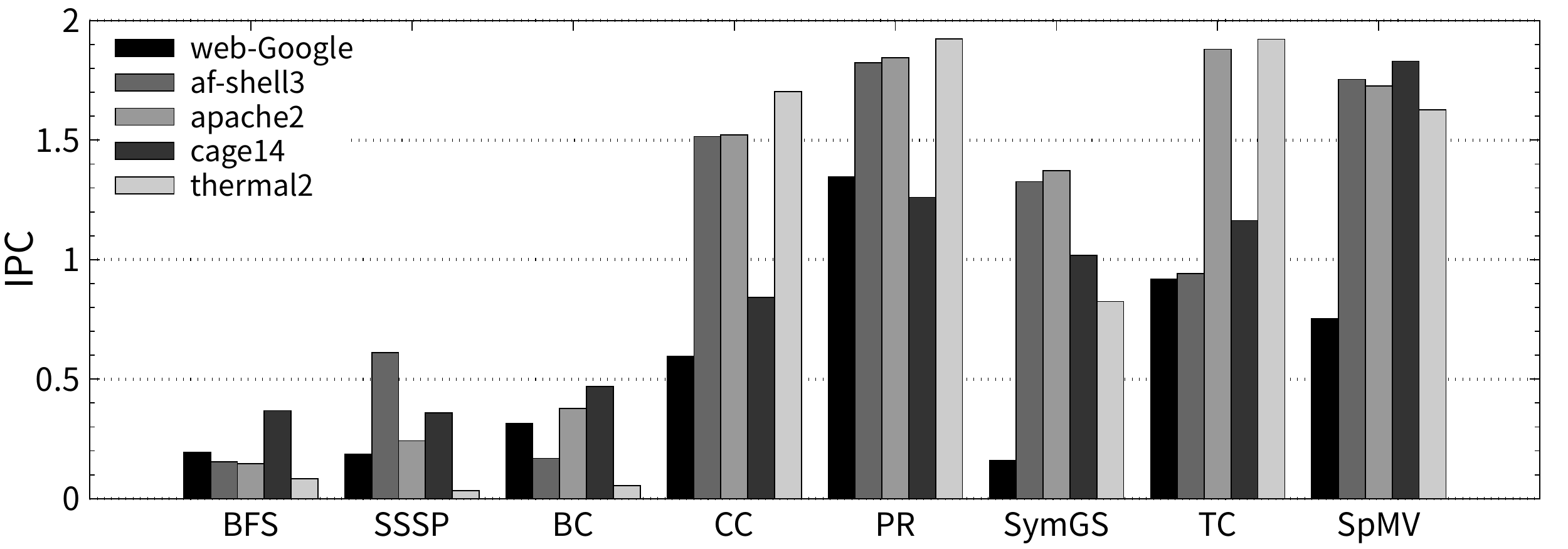}
\vspace{-0.3cm}
\caption{The program IPC for different workloads and datasets.}
\label{fig:diversity_ipc}
\end{figure}

\begin{figure}[!ht]
	\centering
	\includegraphics[width=0.90\textwidth]{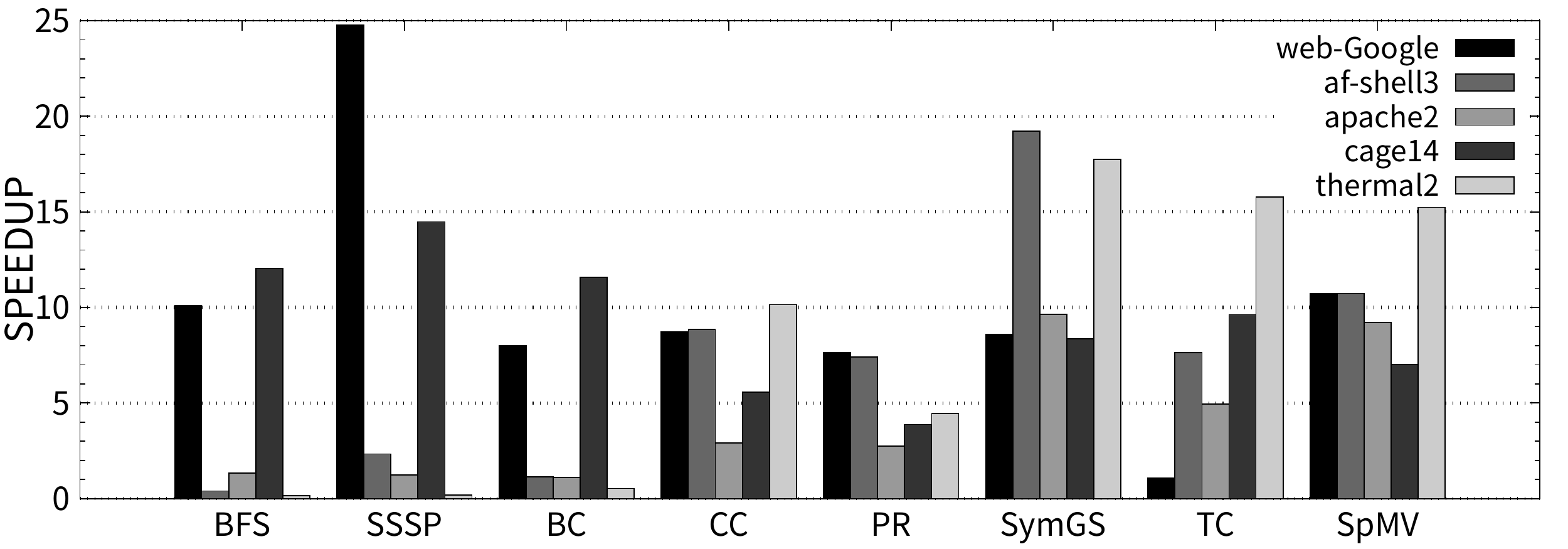}
	\vspace{-0.3cm}
	\caption{The program speedup for different workloads and datasets.}
	\label{fig:diversity_speedup}
\end{figure}

Figure~\ref{fig:diversity_ipc} shows the IPC for 8 workloads running with different input datasets. 
In each workload cluster, IPC varies widely among different datasets. For the same dataset, 
different workloads may lead to entirely different IPCs . For example, for \texttt{thermal2}, 
BFS, SSSP and BC have low IPCs which are below 0.1, while CC, PR, TC and SpMV have much higher 
IPCs (above 1.5). Thus, architecture research needs a diverse set of workloads (in 
terms of application domains and program behaviors) and input datasets (in terms of 
graph size, density, and topology) to make microarchitecture characteristics fully 
manifest, and GARDENIA is designed to meet this requirement.

Figure~\ref{fig:diversity_speedup} shows the speedup of the 8 workloads in comparison with the CPU serial implementations. It shows similar behavior like Figure~\ref{fig:diversity_ipc}. The speedups vary widely with different datasets. For the same dataset, different workloads lead to entirely different speedups. In particular, we can see BFS, SSSP and BC get speedup lower than 1 with some input datasets, which means current CUDA implementations of them are not efficient.




\begin{figure}[!ht]
		\includegraphics[width=0.7\textwidth]{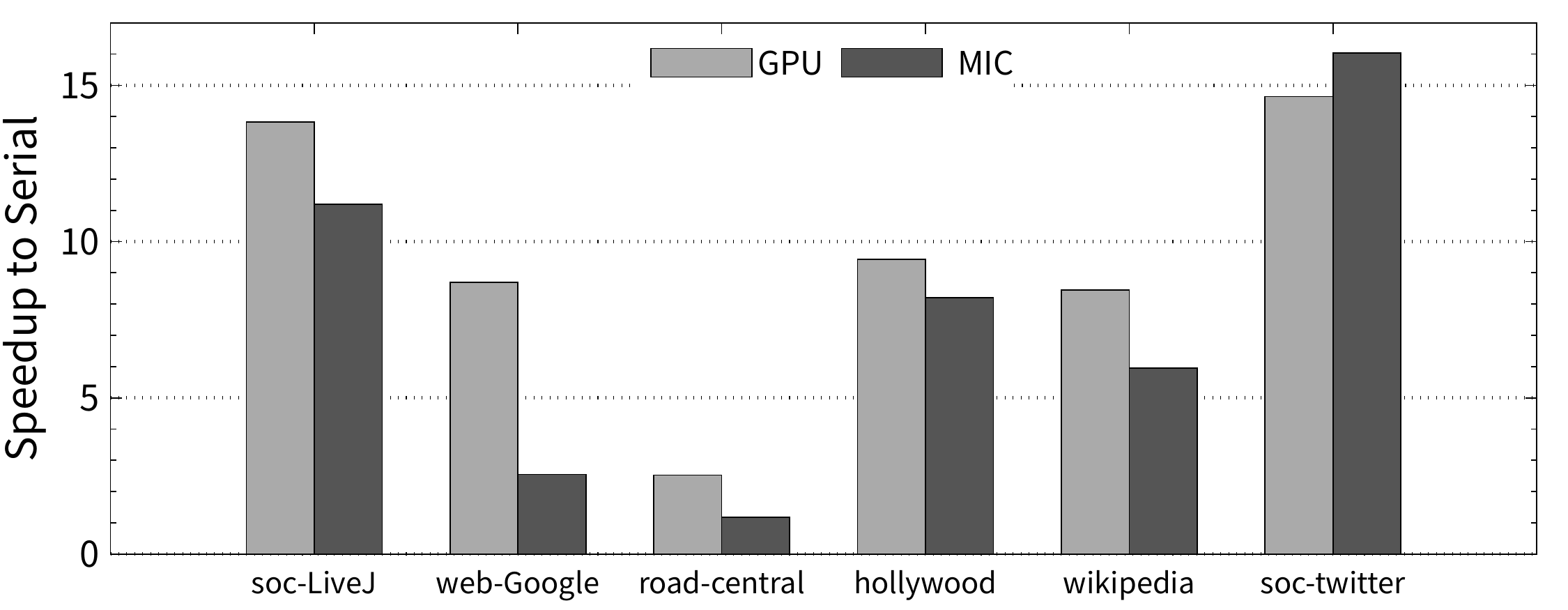}	
	\caption{Speedup comparison between GPU and MIC}
	\label{fig:mic_vs_gpu}
\end{figure}

\subsection{Accelerator Coverage}
GARDENIA targets future accelerator architecture research, including GPU and MIC. It includes both GPU and MIC implementation of all workloads, along with CPU OpenMP implementation. To demonstrate the difference between MIC and GPU, we show their speedups of running BFS on 6 datasets.

Figure \ref{fig:mic_vs_gpu} shows the speedup of GPU and MIC implementations versus the serial CPU implementation of the BFS workload with 6 different datasets. Similar to GPU, graph workloads are also data sensible on MIC. For example, BFS with input \textit{soc-twitter} running on MIC can get speedup of 16.0 $\times$, while the corresponding speedup is only 1.2 $\times$ for input \textit{road\_central}. Due to MIC's lower core number and flexibility, speedup on MIC are relatively lower than that on GPU. 

In fact, due to different features of MIC and GPU, irregular workloads running on them have significantly different microarchitecture behaviors. Thus, it is important to include both GPU and MIC implementations in GARDENIA. We don't discuss in depth the microarchitectural difference details in this paper.

\section{Conclusion}\label{sect:conclusion}
In this paper, we present GARDENIA, a domain-specific benchmark suite for 
studying irregular algorithms on massively parallel accelerators. 
With workloads and datasets carefully selected as well as state-of-the-art
techniques employed, GARDENIA is designed to represent emerging 
real-world applications on modern datacenters, and to facilitate accelerator 
architecture research.

Our characterization illustrates that GARDENIA 
benchmarks exhibit quite different microarchitectural behavior from 
structured workloads and straightforward-implemented graph benchmarks,
and thus demonstrating the necessity of constructing such a suite.


\bibliographystyle{ACM-Reference-Format}
\bibliography{references}
\end{document}